\documentclass[showpacs,preprintnumbers]{revtex4}
\usepackage{amsfonts}
\usepackage{amssymb,epsfig}
\usepackage{amsmath}
\usepackage{graphicx}
\usepackage{array}
\usepackage{subfig}
\allowdisplaybreaks
\begin{document}
\title{\bf Bounds on Higher Derivative $f(R,\Box R,T)$ Models from Energy Conditions}

\author{M. Ilyas}
\email{ilyas_mia@yahoo.com}
\affiliation{Centre for High Energy Physics, University of the Punjab, Quaid-i-Azam Campus, Lahore-54590, Pakistan}

\author{Z. Yousaf}
\email{zeeshan.math@pu.edu.pk}
\affiliation{Department of Mathematics, University of the Punjab, Quaid-i-Azam Campus, Lahore-54590, Pakistan.}

\author{M. Z. Bhatti}
\email{mzaeem.math@pu.edu.pk}
\affiliation{Department of Mathematics, University of the Punjab, Quaid-i-Azam Campus, Lahore-54590, Pakistan.}
\begin{abstract}
This paper studies the viable regions of some cosmic models in a higher derivative $f(R,\Box R, T)$
theory with the help of energy conditions (where $R,~\Box $ and
$T$ are the Ricci scalar, d'Alembert's operator and trace of energy
momentum tensor, respectively). For this purpose, we assume a flat
Friedmann-Lema\^{i}tre-Robertson-Walker metric which is assumed to be filled with perfect
fluid configurations. We take two distinct realistic models, that might be helpful to
explore stable regimes of cosmological solutions. After taking some numerical values of
cosmic parameters, like crackle, snap, jerk (etc) as well as viable constraints from energy conditions,
the viable zones for the under observed $f(R,\Box R, T)$ models are examined.
\end{abstract}

\maketitle

\section{Introduction}

It could be helpful to understand the restrictions on general relativity (GR)
and improve our theoretical predictions on cosmological scales as
modification in GR could provide outstanding results at large scales.
Many relativists used modified gravity theories, after the cosmic accelerating
picture made by the BICEP2 experiment ~\cite{Ade:2014xna, Ade:2015tva, Array:2015xqh},
Wilkinson microwave anisotropy probe \cite{Komatsu:2010fb, Hinshaw:2012aka} and the Planck satellite ~\cite{ya2, Planck:2015xua, Ade:2015lrj}
which illustrate enigmatic forces behind the cosmic evolution.
The dark energy is attributed as an active candidate which influence acceleration in the cosmic expansion.
Qadir et al. \cite{zs3} proposed that GR modification could provide fruitful insights to study issues linked with
quantum gravity and dark matter problem.
Various modified gravity theories have been proposed by relativists which gained significance due to their additional
degrees of freedom (please see modified gravity and DE reviews~\cite{R1,R2,R3,R4,R5,R6,R7,R8,R9,R10,3,4,5,6,8}).

The dark side of the universe could be studied by modified theory of gravity that work on the geometrical side of field equations which includes the study of dark energy and dark matter. Different theories have been suggested associated with higher powers
of Riemann tensor in Lagrangian, that has been demonstrated useful
in cosmology. The most studied framework in this context are $f(R)$ theories, however some suitable modifications are also possible by considering higher derivative terms of the curvature related objects and henceforth called the higher order gravity theories. Starobinsky \cite{a} and Kerner \cite{b} did the pioneering work of introducing the higher derivative terms while
exploring solutions avoiding the initial singularity. Further, it is
established that these higher derivative gravity models could have
significant role in order to study the inflationary expansion of the universe \cite{c}.

Bonanno \cite{d} explored exact solutions of electrically charged spherical interiors containing higher derivative terms
with null fluid. He also discussed the stability of Cauchy horizon and claimed that in the background of anti de Sitter model, the stable solutions are black holes. Cuzinatto et al. \cite{e} studied the
higher derivative theory in which the Lagrangian involve terms of order $n$, e.g.,
$f(R, \nabla_\mu R,\nabla_{\mu1}\nabla_{\mu2}R,...,\nabla_{\mu1}...\nabla_{\mu \textmd{n}}R)$,
and performed transformation from Jordan-to-Einstein frame in both metric and Palatini formalism.
Iihoshi \cite{f} presented a hybrid inflationary scenario in the background of
$f(R,\Box R)$ theory, where $\Box$ indicate the d'Alembertian operator. This theory is treated as instinctive
generalization of $f(R)$ greavity. Also, he proposed that $f(R,\Box R)$ theory is equivalent to GR coupled with two scalar fields.

By using the dynamical system technique, Tretyakov \cite{g}
presented the Minkowski stability issue in the scenario of particular modified theory, i.e. $f(R)+R\Box R$ gravity.
He claimed that this method is useful for extracting additional constraints on parameters of various modified gravity theories.
There also exists a well-known class of gravity theories in which a
more general function of Ricci scalar $R$ replaces the arbitrary
function $f(R)$ in the gravitational action of GR.
One of such theories is $f(R,T)$ gravity, firstly discussed by Harko
\textit{et al.} \cite{16} in which the arbitrary function,
incorporates the energy-momentum trace $T$ along with $R$. They also
analyzed the self-interacting scalar field models and the Newtonian
limit of these modified models. Recently, Abbas \emph{et al.} \cite{abbas} have discussed the viability of modified
gravity models through gravitational collapse.

Initially, Houndjo \emph{et al.} \cite{26}
discussed the FLRW cosmology in the context of $f(R,\Box R, T)$
gravity and found unstable phase of de Sitter model in this scenario. Yousaf \emph{et al.} \cite{18}
discussed the energy conditions (ECs) and the behavior of Friedmann-Lema\^{i}tre-Robertson-Walker (FLRW) model in
$f(R,\Box R, T)$ gravity. They also showed the constraints and the graphical
behavior of some of the model parameters. Alvarenga \textit{et al.}
\cite{19} investigated the matter density perturbations in modified
$f(R,T)$ models of type $f_{1}(R)+f_{2}(T)$, satisfying
stress-energy conservation and also compared the results of
quasi-static approximations in $f(R,T)$ with that of GR results.
They concluded that the unusual behavior of the density contrast
constrains the viability of such modified models.

Baffou \textit{et
al.} \cite{20} focused on the cosmological dynamics of low and high
red-shift solutions and stability of a modified model, $R+f(T)$, by
using the power law and de Sitter solution with linear perturbation
and concluded the viability of the considered $f(R,T)$ models.
Ilyas \textit{et al.} \cite{24} explored the formation of compact
structures with anisotropic matter content in modified $f(R,T)$
background and concluded the maximum value of pressure and density
in the central region. The spherical hydrostatic equilibrium
configuration of stellar remnants in the background of
$f(R,T)=R+\lambda T$ is analyzed by Moraes \textit{et al.} \cite{25}. Sahoo \emph{et al.} \cite{sahoo} also found
some interesting results based on $f(R,T)$ gravity.
Myrzakulov \cite{17} geometrically constructed few
$f(R,\mathcal{T})$ models (where $\mathcal{T}$ is the torsion scalar) including models of the form, $f(R,\mathcal{T})=\mu R+\nu \mathcal{T}$ and claimed
that some of the cosmological outcomes of this theory could explain the phenomenon of accelerated
expansion of the universe.

A worth emphasizing aspect in the discussion of singularity theorems and black hole thermodynamics is the ECs
which were initially derived in GR by Hawking and Ellis \cite{27}.
Several cosmological issues like, expansion history of the universe, phantom fields etc, have been
discussed by using the ECs in GR.
Also, one can explore additional constraints by exploring EC in the analysis of modified theories to derive general
results that hold for a variety of situations. Santos \emph{et al.} \cite{28} studied the bounds enforced by ECs in the context of $f(R)$
functional form. They explored the null, strong, weak and dominant ECs from Raychaudhuri equation
which differ from those obtained in GR.
Bertolami and Sequeira \cite{29} derived the ECs
for a particular type of gravity model having non-minimal matter-curvature coupling
and discussed their stability via Dolgov-Kawasaki criterion. Zhou \emph{et al.} \cite{ad1} after considering FLRW metric proposed two stable $f(G)$
toy models (where $G$ is the Gauss-Bonnet term) for discussing a phantom-like and de Sitter environment. Atazadeh and Darabi \cite{ad2} assumed two different formulations of $f(R,G)$ gravity and developed viable bounds through ECs.

This paper is aimed to prob the issue of viability of $f(R,\Box R,T)$ models in an environment of FLRW perfect fluid
metric. Our work is organized as under. The next section will briefly describe $f(R,\Box R, T)$ theory along with the
corresponding ECs. Section 3 describes viability constraints coming from ECs for two
$f(R,\Box R, T)$ cosmic models. In the last section, we summarize our main findings.

\section{$f(R,\Box R, T)$ gravity and Energy Conditions}

The usual Einstein Hilbert action (EHA) for $f(R,\Box R, T)$ theory can be modified as under \cite{26}
\begin{equation}\label{action}
S = \frac{1}{2\kappa^2}\int {{d^4}x\sqrt { - g} f(R,\Box R, T) + {S_M}\left( {{g^{\mu \nu }},\psi } \right)},
\end{equation}
where $T$ and $R$ are the traces of the stress-energy
and Ricci tensors, respectively, while ${\kappa ^2} = 8\pi G$ with
$G$ as the Newton's gravitational constant. The operator $\Box\equiv\nabla_\mu\nabla^\mu$, in which
$\nabla_\mu$ describes covariant derivation. Variations of the above modified EHA with the metric tensor provides
\begin{equation}\label{action1}
\delta S = \frac{1}{2\kappa^2}\int {d^4}x[f\delta \sqrt{-g}+\sqrt{-g}(f_R\delta R+f_{\Box R}\delta\Box R+f_T\delta T)
+2\kappa^2 \delta(\sqrt{-g}L_M)],
\end{equation}
where $L_M$ indicates Lagrangian for matter field. Equation \eqref{action1} after substituting $\delta
R,~\delta\Box R,~\delta \sqrt{-g}$ and $\delta T$ with few calculations gives
\begin{align}\nonumber
\delta S &= \frac{1}{2\kappa^2}\int {d^4}x\left[-\frac{1}{2}\sqrt{-g}g_{\alpha\beta}\delta g^{\alpha\beta}f
+\sqrt{-g}(T_{\alpha\beta}+\Theta_{\alpha\beta})f_T\delta g^{\alpha\beta}
+f_R\sqrt{-g}(R_{\alpha\beta}\right.\\\nonumber
&\left.+g_{\alpha\beta}\Box-\nabla_\alpha\nabla_\beta)\delta g^{\alpha\beta}
+\sqrt{-g}f_{\Box R}(\nabla_\alpha\nabla_\beta R+\Box R_{\alpha\beta}+g_{\alpha\beta}\Box^2
+R_{\alpha\beta}\Box-\Box \right.\\\label{action2}
&\left.\times\nabla_\alpha\nabla_\beta-\nabla_\alpha R\nabla_\beta+2g^{\mu\nu}\nabla_\mu
R_{\alpha\beta}\nabla_\nu)\delta g^{\alpha\beta}+2\kappa^2\frac{\delta(\sqrt{-g}L_M)}{\delta g^{\alpha\beta}}\delta
g^{\alpha\beta}\right],
\end{align}
where subscripts $R,~T$ and $\Box R$ indicate the derivative of the
corresponding quantities with respect to $T,~R$ and $\Box R$,
respectively, while $\Theta_{\alpha\beta}=g^{\mu\nu}\delta
T_{\mu\nu}/\delta g^{\alpha\beta}$. Equation (\ref{action2}) after
simplifications gives rise to
\begin{equation}\label{fieldeq}
\begin{gathered}
{{f_R}{R_{\alpha\beta}}} + \left( {{g_{\alpha\beta}}\Box  - {\nabla _\alpha }{\nabla _\beta }} \right){{f_R}}
- \frac{1}{2}{g_{\alpha\beta}}f + \left( {2{f_{\Box R}}
({\nabla _{(\alpha }}{\nabla _{\beta )}})} \right.R\left. { - \Box {R_{\alpha\beta}}} \right)\\
- \left\{ {{R_{\alpha\beta}}}\Box - \Box {\nabla _\alpha }{\nabla _\beta }
+ {g_{\alpha\beta}}{\Box ^2} - {\nabla _\alpha }R{\nabla _\beta }
+2 {g^{\mu\nu }}{\nabla _\mu } {{R_{\alpha\beta}}{\nabla _\nu }} \right\}{f_{\Box R}}\\
=\kappa^2 T_{\alpha\beta}-f_T (T_{\alpha\beta}+\Theta_{\alpha\beta}),
\end{gathered}
\end{equation}
where $\Theta_{\alpha\beta}$, after selecting $L_m=-p$, turns out to be ${\Theta_{\alpha\beta}=2
T_{\alpha\beta }-p g_{\alpha\beta}}$.

We start our analysis by taking a FLRW spacetime
which in the background of flat homogeneous state can be given as follows
\begin{equation}\label{metric}
d{s^2} =  - d{t^2} + {a^2}(t)(d{x^2} + d{y^2} + d{z^2}),
\end{equation}
in which $a$ is the scale factor. The stress-energy tensor for the perfect fluid describes
the contributions of pressure $(p)$ and energy density $(\rho)$. This in terms of mathematical expression can be given as
\begin{equation}\label{emtensor}
T_{\alpha\beta}=(\rho+p)u_\alpha u_\beta -p g_{\alpha\beta}.
\end{equation}
Here, the fluid four velocity is given by $u_\mu$. The
corresponding $f(R,\Box R, T)$ equations of motion (\ref{fieldeq})
for the FLRW spacetime (\ref{metric}) and
(\ref{emtensor}) are \cite{26}
\begin{align}\nonumber
&2H{f_{\Box R}}^{\prime \prime \prime} - \left( {2{H^2} + 3H'}
\right){f_{\Box R}}^{\prime \prime } - (5{H^3} + 2HH'
+ {H^{\prime \prime }}){f_{\Box R}}^\prime + 2\{-2{H^2}H' + 6{{H'}^2} \\\label{tf1}
& + 3H{H^{\prime \prime }} + H^{\prime \prime \prime } \}
{f_{\Box R}} + H{f_R}^\prime - \left( {{H^2} + H'} \right){f_R}
- \frac{f}{6} = \frac{1}{3}[\rho {\kappa ^2} + (\rho  + p){f_T}],\\\nonumber
&{f_{\Box R}}^{\prime \prime \prime \prime} + 5H{f_{\Box R}}^{\prime \prime \prime}
+ \left( { - 8{H^2} + 5H'} \right){f_{\Box R}}^{\prime \prime } + ( - 23{H^3} + 2HH'
+ 4{H^{\prime \prime }}){f_{\Box  R}}^\prime \\\nonumber
&+ 2\left( { - 2{H^2}H' + 6{{H'}^2} + 3H{H^{\prime \prime }} + {H^{\prime \prime \prime }}} \right){f_{\Box R}}
 - 2H{f_R}^\prime - \left( {3{H^2} + H'} \right){f_R}\\\label{tf2}
& -{f_R}^{\prime \prime } - \frac{f}{2} = {\kappa ^2}p.
\end{align}
In terms of the FLRW scale factor, the Hubble $(H)$ and deceleration $(q)$ are calculated as
\begin{align}\label{hq}
H=\frac{\dot{a}}{a},\quad q =-\frac{1}{{{H^2}}}\frac{{a''}}{a},
\end{align}
whereas jerk $(j)$, snap $(s)$ and crackle $(l)$ are
\begin{align}\label{jsl}
j=\frac{{a'''}}{aH^3},\quad s = \frac{{a''''}}{aH^4}\quad
l=-\frac{{a'''''}}{aH^5}.
\end{align}

The study of the exploring viable bounds on gravity models through ECs has
been a source of great interest by many mathematical physicists. The constraints obtained through ECs could lead to analyze the
stability of some relativistic systems. The coupling of fluid distributions with the
various geometries, like spherical, axial, cylindrical
symmetries are often in practice during the modeling of compact objects. In order to have
the realistic configurations of these fluids, one must pick the viable formulations of stress-energy tensor.
The idea of ECs could provide an effective tool in this regard, in the sense that only those stress-energy
tensors are realistic, that satisfy the corresponding ECs. These conditions are
observed to be coordinate-invariant (independent of symmetry). In the standpoint of expanding nature,
the Raychaudhuri's equation can be stated as
\begin{align}
\frac{d \Theta_1}{d\tau}=-\frac{\Theta_1}{2}+\omega^{\alpha\beta}\omega_{\alpha\beta}
-\sigma^{\alpha\beta}\sigma_{\alpha\beta}-R_{\alpha\beta}k^\alpha k^\beta,
\end{align}
where $\omega_{\alpha\beta}$ and $\sigma_{\alpha\beta}$ are rotation and shear tensor, respectively,
whereas $\Theta_1$ is an expansion scalar. These quantities are characterized
by the congruences connected with the null vector $k_\mu$.
Bamba \emph{et al.} \cite{ps1} explored few stable bounds through ECs in $f(G)$ gravitational theories.
The following distributions of ECs can be formulated in terms of effective forms of energy density and pressure as
\begin{equation}
NEC  \Leftrightarrow {\rho _{eff}} + {p_{eff}} \ge 0,
\end{equation}
\begin{equation}
WEC \Leftrightarrow {\rho _{eff}} \ge 0 \text{ and } {\rho _{eff}} + {p_{eff}} \ge 0,
\end{equation}
\begin{equation}
SEC \Leftrightarrow {\rho _{eff}+ 3{p_{eff}}} \ge 0 \text{ and } {\rho _{eff}} + {p_{eff}} \ge 0,
\end{equation}
\begin{equation}
DEC \Leftrightarrow {\rho _{eff}} \ge 0 \text{ and } {\rho _{eff}} \pm {p_{eff}} \ge 0.
\end{equation}
We define a parameter $m$ as follows
\begin{equation}
\begin{gathered}
m =  - \frac{1}{{{H^6}}}\frac{{a^{(5)}}}{a}.
\end{gathered}
\end{equation}
In terms of cosmic parameters, the derivatives of Hubble parameter $H$ are found to be
\begin{equation}
\begin{gathered}
H' =  - {H^2}\left( {1 + q} \right),\\
{H^{\prime \prime }} = {H^3}\left( {j + 3q + 2} \right),\\
{H^{\prime \prime \prime }} = {H^4} \left( {s - 4 j - 12 q - 3 q^2- 6} \right),\\
{H^{\prime \prime \prime \prime}} = {H^5}\left( {24 + l + 60q + 30{q^2} + 10j\left( {2 + q} \right) + 5s} \right),\\
{H^{(5)}} = {H^6}\left( { - 10{j^2} - 120j(q + 1) + 6l + m - 30{q^3} - 270{q^2} + 15qs - 360q + 30s - 120} \right).
\end{gathered}
\end{equation}
In order to solve cumbersome and lengthy $f(R,\Box R,T)$ equations of motion, we consider $f_T=0$. In this framework, the 00 field
equation provides
\begin{equation}\label{ro}
\begin{gathered}
\rho  =  - \frac{f}{2} + f_{\Box R}^{\prime \prime \prime } + 3(H(H(3q + 1)f_{\Box R}^{\prime \prime } + {H^2}f_{\Box R}^\prime ( - (j + q + 5))
 + f_R^\prime  - f_R H q) + 2f_{\Box R}(H''' + {H^4}(3j + 6{q^2} + 23q + 14))),
\end{gathered}
\end{equation}
while the sum of energy density and pressure gives
\begin{equation}\label{ropp}
\begin{gathered}
\rho  + p = {f_{\Box R}}^{\prime \prime \prime \prime } - 2f_R{H^2} + 112{f_{\Box R}}{H^4} + {f_{\Box R}}^{\prime \prime \prime}\left( {1 + 5H} \right) + 8{f_{\Box R}}{H^{\prime \prime \prime }} + 24{f_{\Box R}}{H^4}j - 2f_R{H^2}q\\
 + 184{f_{\Box R}}{H^4}q + 48{f_{\Box R}}{H^4}{q^2}
 - 44{H^3}{f_{\Box R}}^{\prime} - 7{H^3}j{f_{\Box R}}^{\prime} - 13{H^3}q {f_{\Box R}}^{\prime}
 + H {f_{R}}^{\prime} + 16{H^2}{f_{\Box R}}^{\prime \prime}\\+ 14{H^2}q {f_{\Box R}}^{\prime \prime} - f_{R}^{\prime \prime }.
\end{gathered}
\end{equation}

\section{Different Models}

The purpose of this work is to present some viable regions of $f(R,\Box R,T)$ models. We want to analyze the behavior of
ECs for the perfect and flat FLRW metric by considering the case of separating $R$ and $\Box R$ formulations. Here, we take
models of the forms
$$f(R,\Box R)=f(R)+f(\Box R).$$
We proceed our work by considering the following choices of cosmological parameters
\begin{equation}
\begin{gathered}\label{para1}
H = 0.718, q =  - 0.64, j = 1.02,\\
s =  - 0.39, l = 3.22, m=-11.5.
\end{gathered}
\end{equation}
We shall use above values as well as separable combinations of $f(R)$ and $f(\Box R)$ functions to explore
ECs in the following subsections.

\subsection{Model 1}

\begin{figure} \centering
\epsfig{file=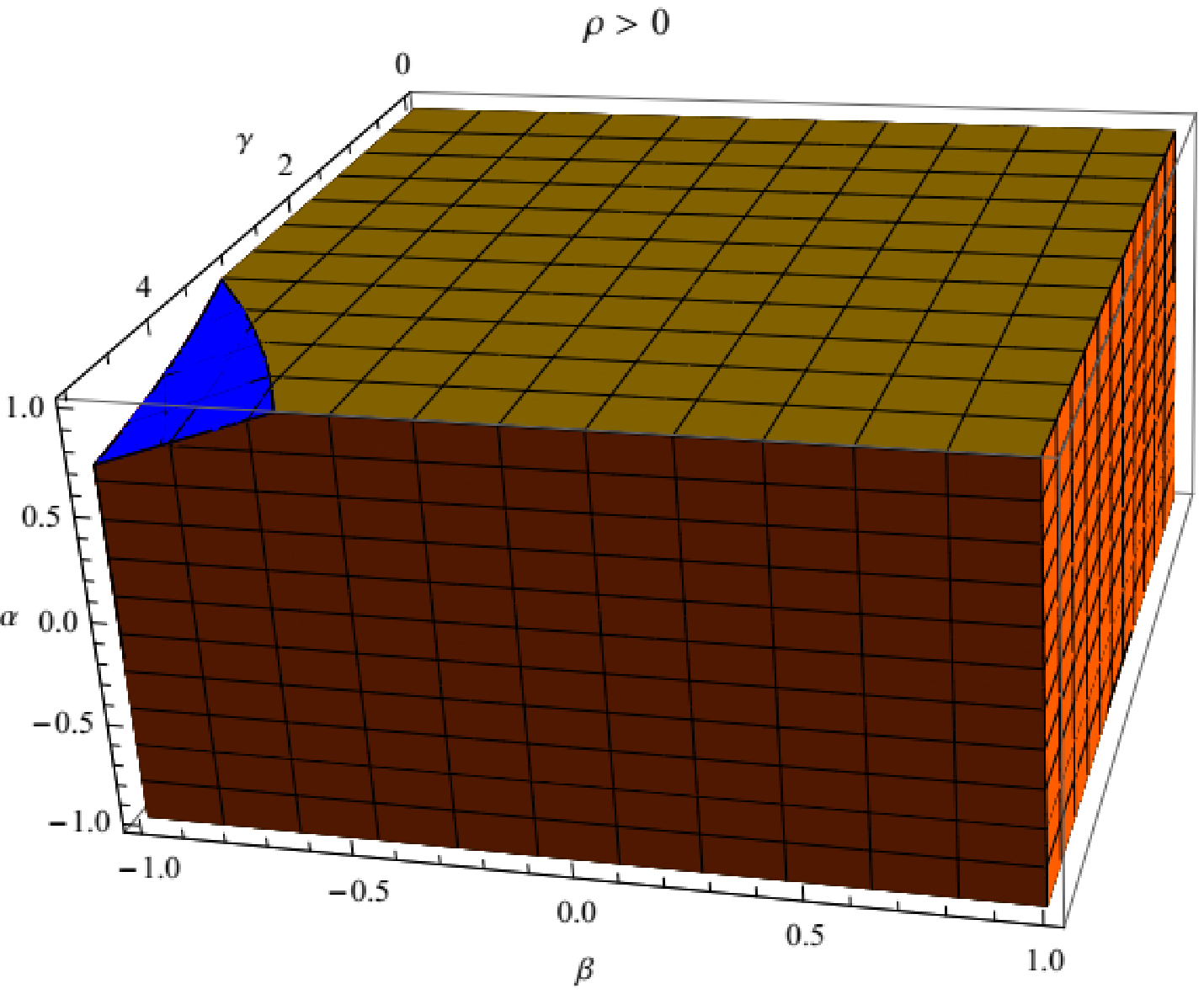,width=.48\linewidth}
\epsfig{file=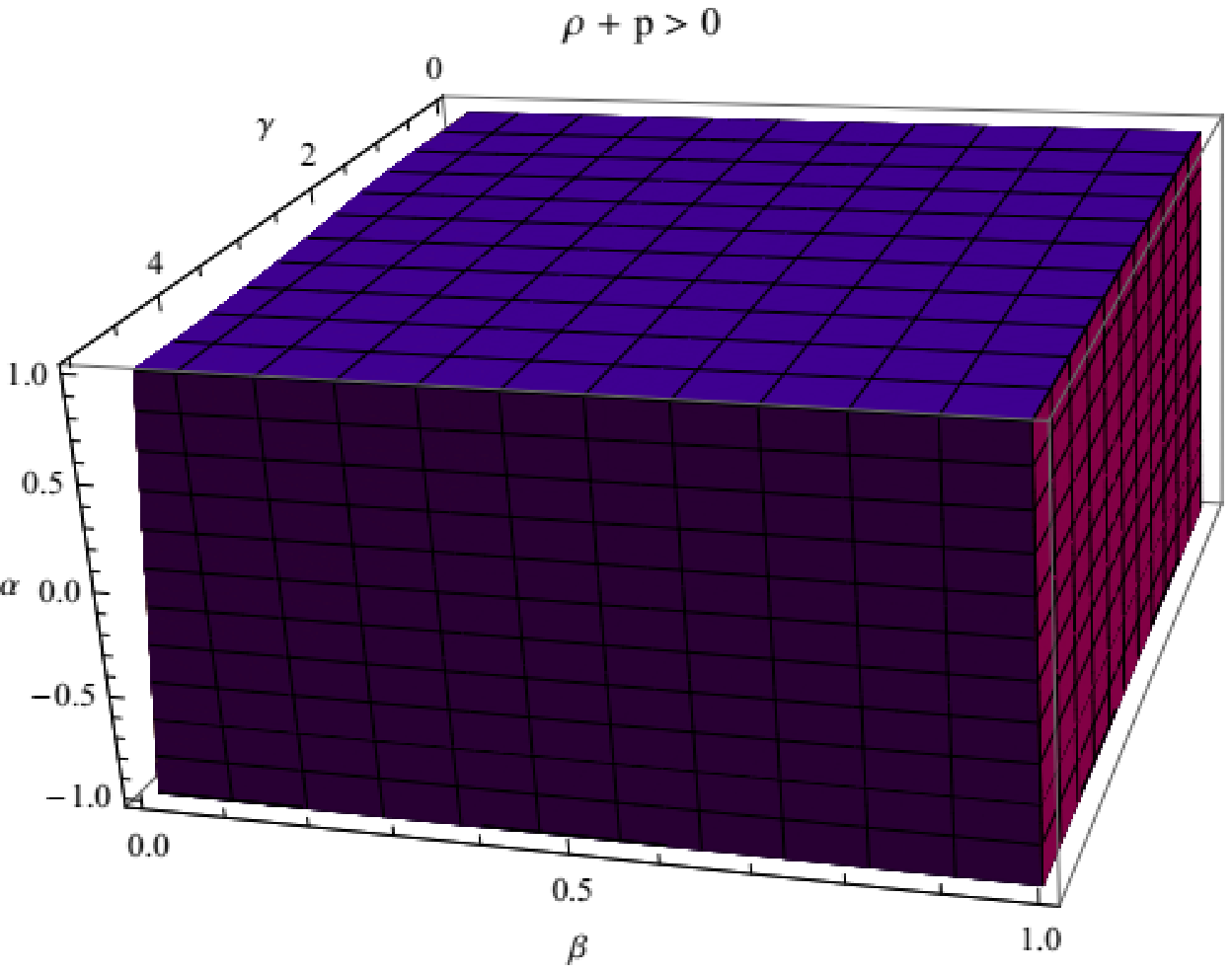,width=.48\linewidth}
\caption{Plot of WEC for model 1 as in Eq.(\ref{model1}), left plot show $\rho$ while the right plot show the $\rho+ p$ with respect to $\alpha$, $\gamma$ and $\beta$}\label{f11}
\end{figure}

\begin{figure} \centering
\epsfig{file=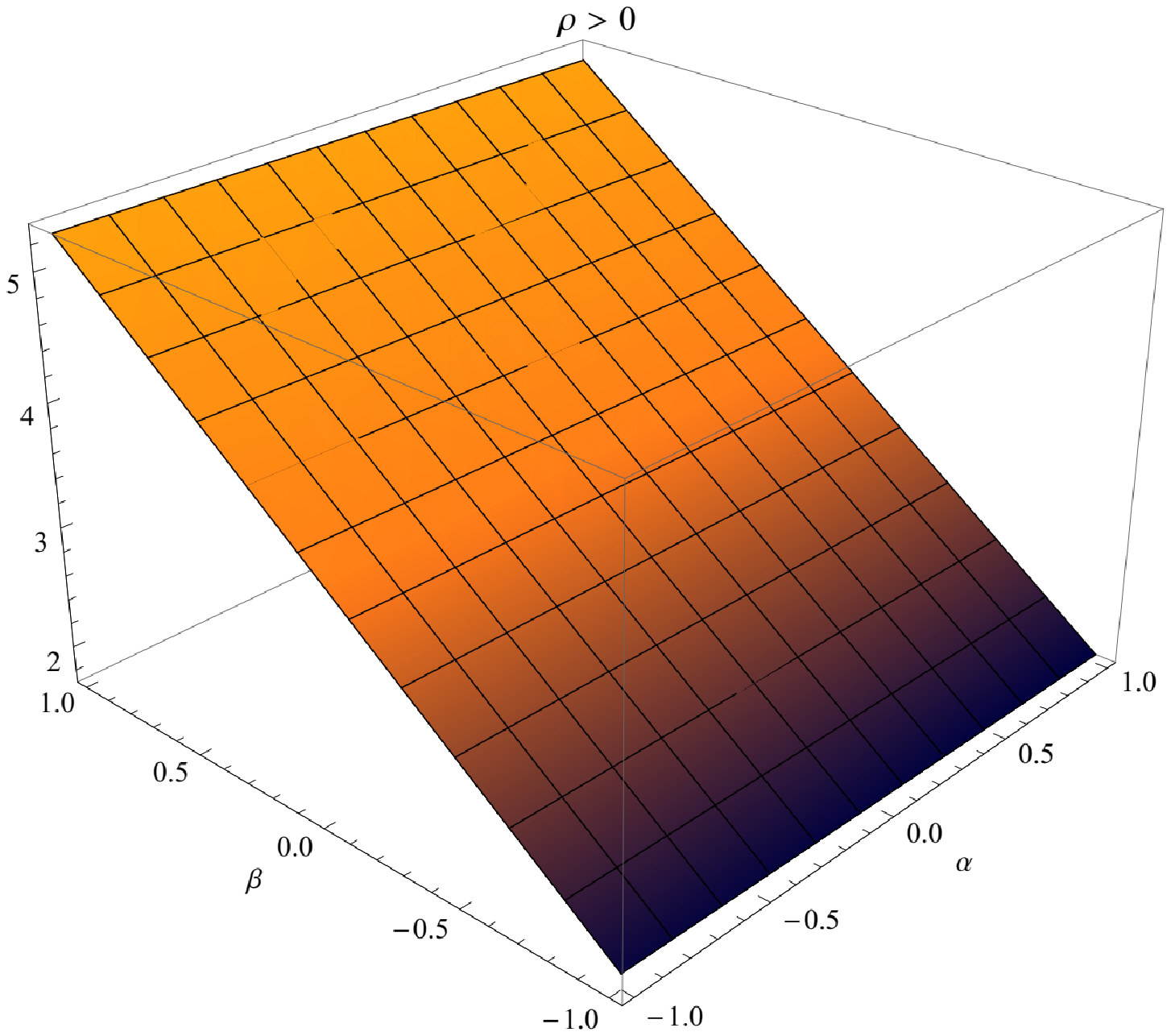,width=.48\linewidth}
\epsfig{file=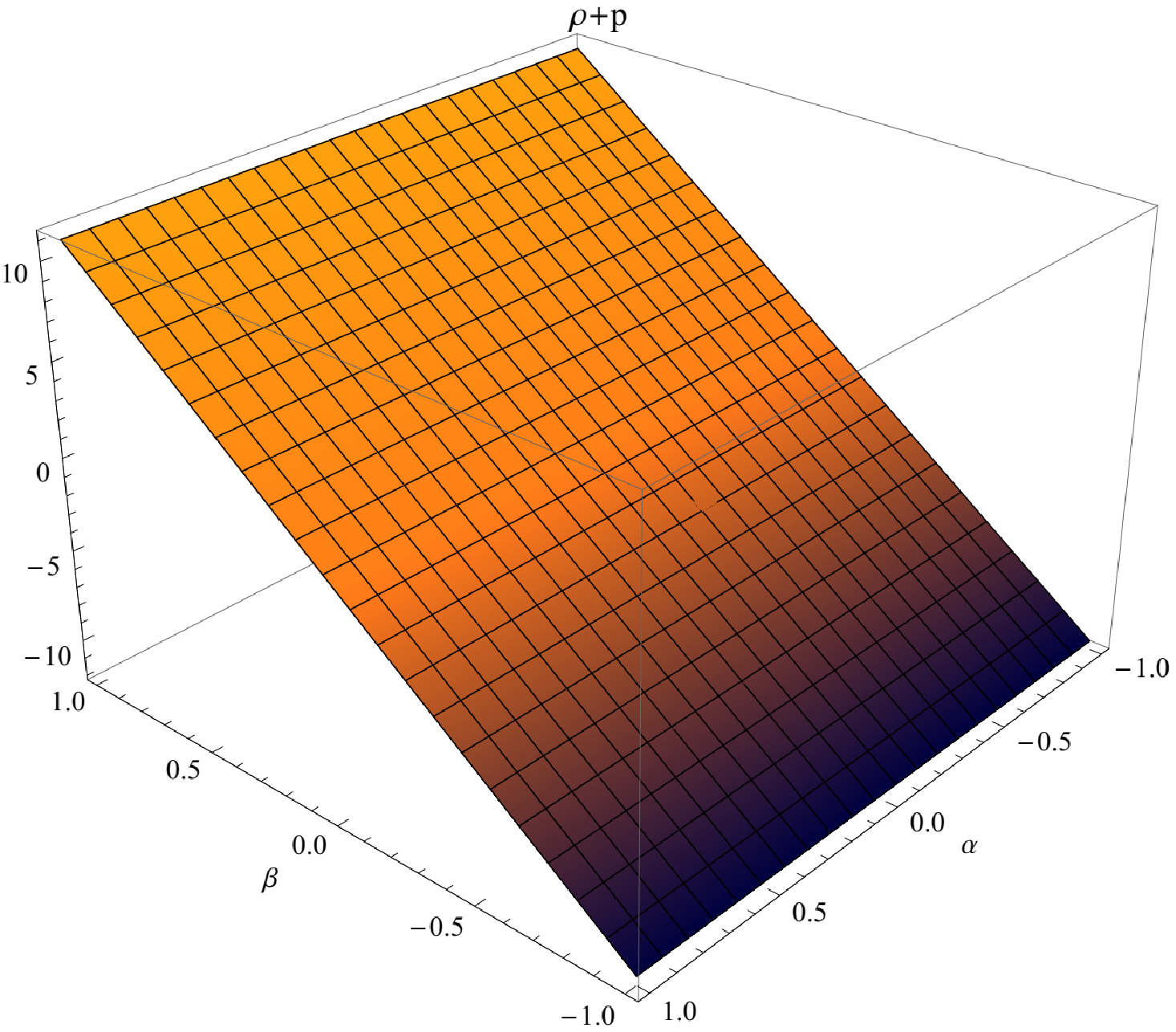,width=.48\linewidth}
\caption{Plot of WEC for model 1 as in Eq.(\ref{model1}), left plot show $\rho$ while the right plot show the $\rho+ p$ with respect to $\alpha$ and $\beta$}\label{f12}
\end{figure}

\begin{figure} \centering
\epsfig{file=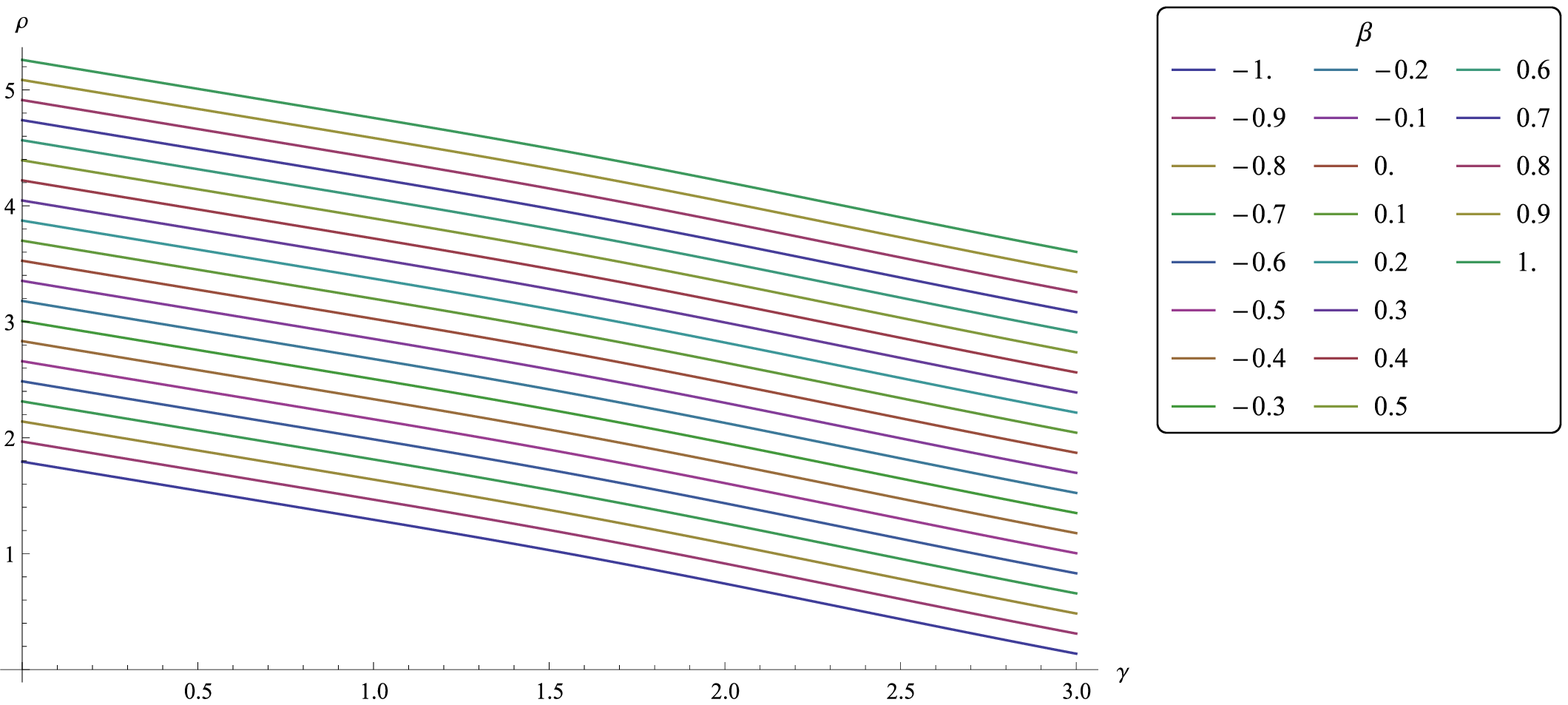,width=.48\linewidth}
\epsfig{file=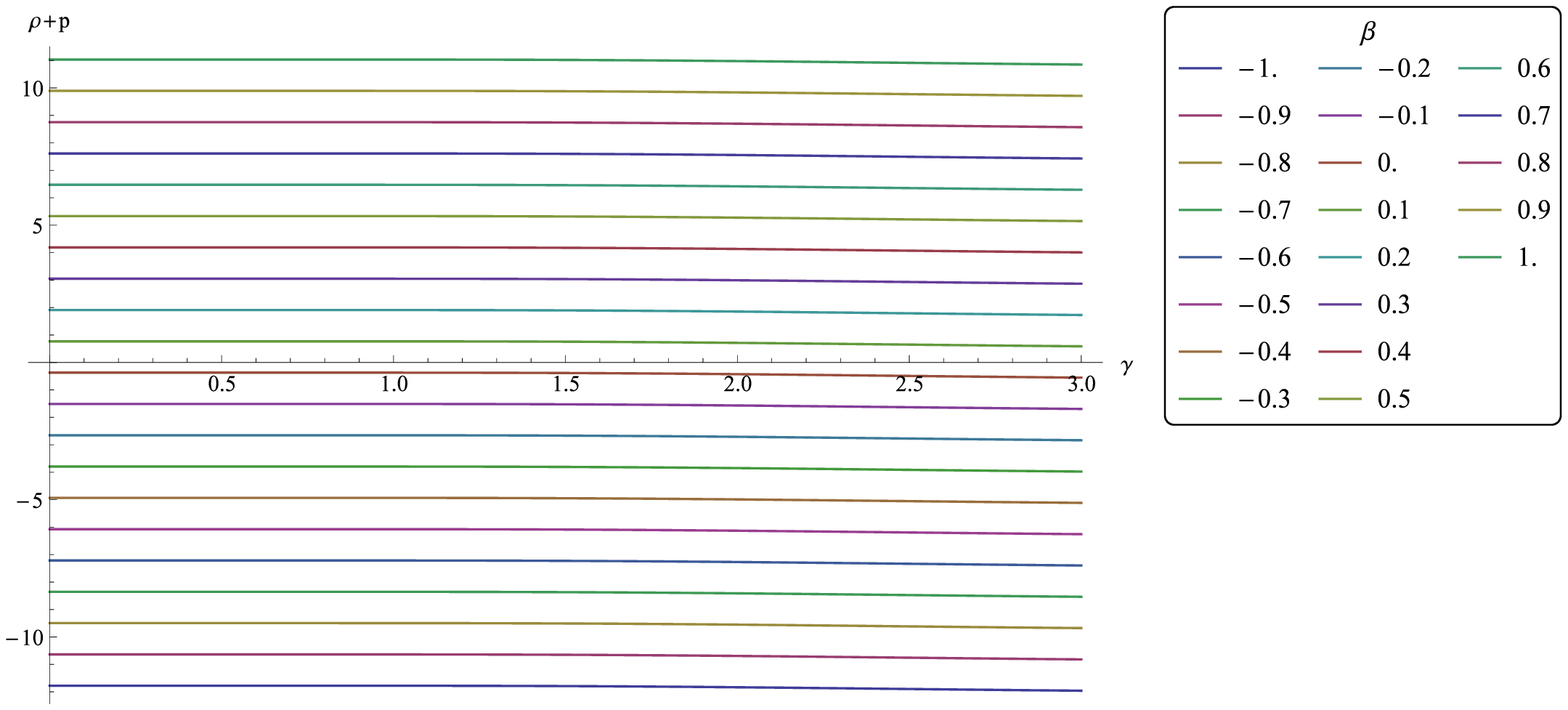,width=.48\linewidth}
\caption{Plot of WEC for model 1 as in Eq.(\ref{model1}), left plot show $\rho$ while the right plot show the $\rho+ p$ with respect to $\gamma$ having $\alpha>0$. }\label{f13}
\end{figure}

\begin{figure} \centering
\epsfig{file=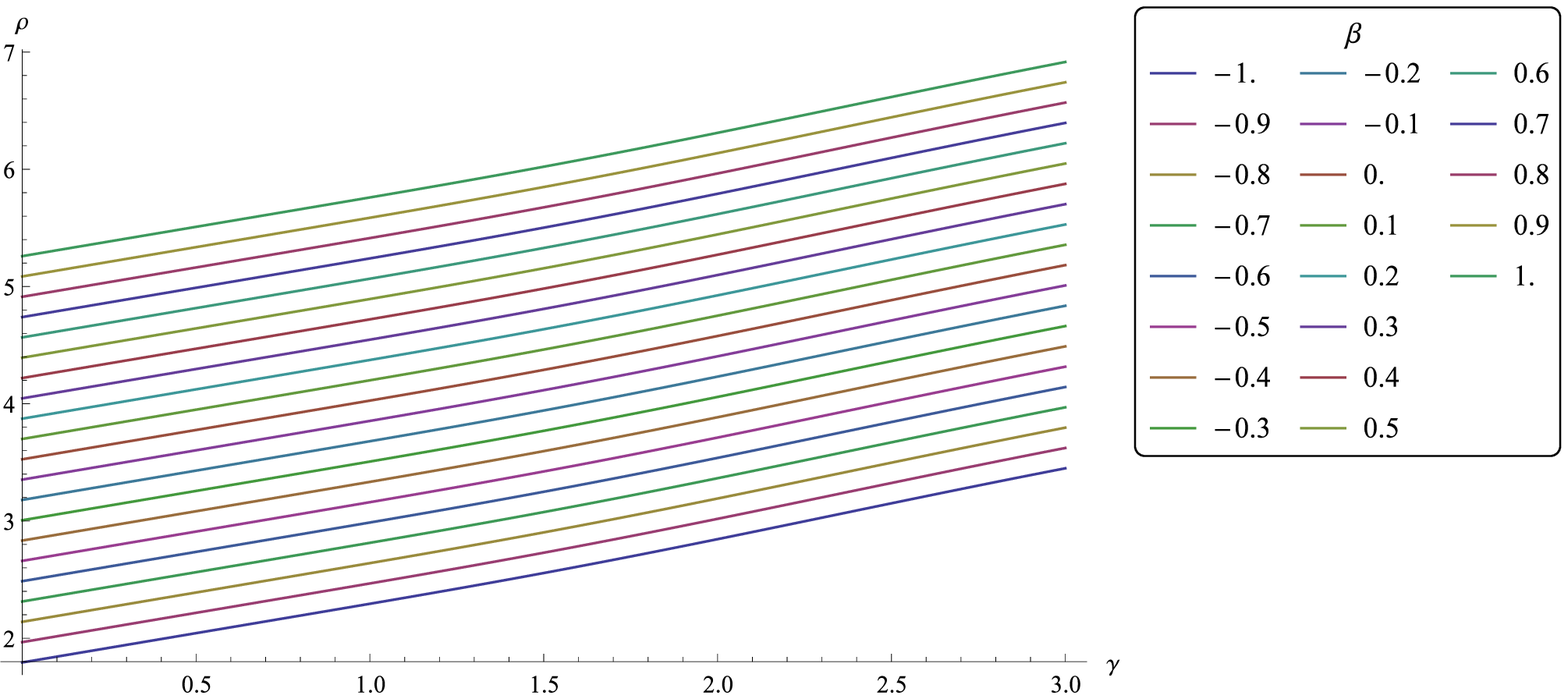,width=.48\linewidth}
\epsfig{file=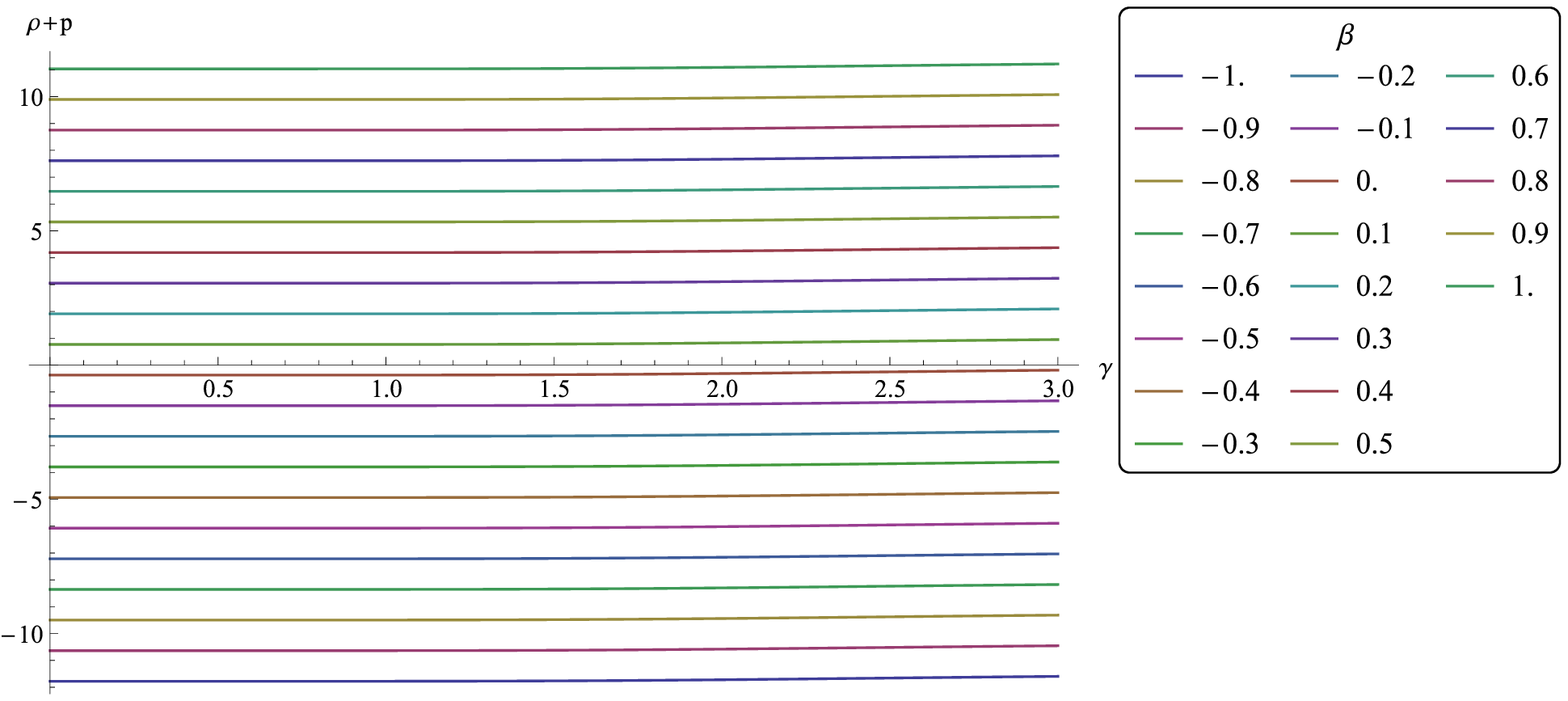,width=.48\linewidth}
\caption{Plot of WEC for model 1 as in Eq.(\ref{model1}), left plot show $\rho$ while the right plot show the $\rho+ p$ with respect to $\gamma$ having $\alpha<0$.}\label{f14}
\end{figure}

First, we choose the tanh Ricci scalar function along with $\beta R \Box R$ term as follows
\begin{equation}\label{model1}
f(R,\Box R) = R - \alpha \gamma \tanh \left( {\frac{R}{\gamma }} \right) + \beta R \Box R,
\end{equation}
where $\alpha,~\beta$ and $\gamma$ are constants. After using the above $f(R,\Box R)$ model along with the values of cosmological parameters from Eq.\eqref{para1}, Eq.(\ref{ro}) becomes
\begin{align}\nonumber
\rho  &= \frac{1}{{2\gamma }}\left[12\beta \gamma {H^5}\left( {24 - l + 48q + 18{q^2} + 2j\left( {4 + q} \right) - s} \right) + 36\beta \gamma {H^6}(4 + {j^2}
 - l + j\left( {13 - 9q} \right) + 55q\right.\\\nonumber &\left.+ 35{q^2} + 5{q^3} - qs) + 6\gamma {H^2}\left[1
 + q\left\{ { - 2 + \alpha Sech{{\left( {\frac{{6{H^2}(- 1 + q)}}{\gamma }} \right)}^2}} \right\}\right] + \alpha {\gamma ^2}\tanh \left[ {\frac{{6{H^2}\left( { - 1 + q} \right)}}{\gamma }} \right] \right.\\\label{g1}
 &\left.+ 72\alpha {H^4}(2- j + q)Sech{\left[ {\frac{{6{H^2}\left( { - 1 + q} \right)}}{\gamma }} \right]^2}\tanh \left[ {\frac{{6{H^2}\left( { - 1 + q} \right)}}{\gamma }} \right]\right],
\end{align}
whereas Eq.(\ref{ropp}) turns out to be
\begin{align}
\rho  + p &= \frac{1}{{{\gamma ^2}}}{H^2}[6\beta {\gamma ^2}{H^3}(24 - l + 48q + 18{q^2} + 2j\left( {4 + q} \right) - s
) - {\gamma ^2}\left( {1 - 2\alpha  + \cosh \left\{ {\frac{{12{H^2}\left( { - 1 + q} \right)}}{\gamma }} \right\}} \right)\\\nonumber
&\times(1 + q)Sech{\left[ {\frac{{6{H^2}\left( { - 1 + q} \right)}}{\gamma }} \right]^2}
 - 12\alpha \gamma {H^2}\left( { - 8 + j - 9q - {q^2} + s} \right)Sech{\left[ {\frac{{6{H^2}\left( { - 1 + q} \right)}}{\gamma }} \right]^2}\tanh \left[ {\frac{{6{H^2}\left( { - 1 + q} \right)}}{\gamma }} \right]\\\nonumber
& + 6{H^4}\{64\beta {\gamma ^2} - 9\beta {\gamma ^2}l + 313\beta {\gamma ^2}q + 178\beta {\gamma ^2}{q^2} + 18\beta {\gamma ^2}{q^3} + 4\beta {\gamma ^2}s - 2\beta {\gamma ^2}qs - 48\alpha Sech{\left[ {\frac{{6{H^2}\left( { - 1 + q} \right)}}{\gamma }} \right]^4}\\\nonumber
& - 48\alpha qSech{\left[ {\frac{{6{H^2}\left( { - 1 + q} \right)}}{\gamma }} \right]^4}
 - 12\alpha {q^2}Sech{\left[ {\frac{{6{H^2}\left( { - 1 + q} \right)}}{\gamma }} \right]^4} + 96\alpha Sech{\left[ {\frac{{6{H^2}\left( { - 1 + q} \right)}}{\gamma }} \right]^2}\tanh {\left[ {\frac{{6{H^2}\left( { - 1 + q} \right)}}{\gamma }} \right]^2}\\\nonumber
& + 96\alpha qSech{\left[ {\frac{{6{H^2}\left( { - 1 + q} \right)}}{\gamma }} \right]^2}\tanh {\left[ {\frac{{6{H^2}\left( { - 1 + q} \right)}}{\gamma }} \right]^2} + 24\alpha {q^2}Sech{\left[ {\frac{{6{H^2}\left( { - 1 + q} \right)}}{\gamma }} \right]^2}\tanh {\left[ {\frac{{6{H^2}\left( { - 1 + q} \right)}}{\gamma }} \right]^2}\\\nonumber
& + 4{j^2}(\beta {\gamma ^2} - 3\alpha Sech{\left[ {\frac{{6{H^2}\left( { - 1 + q} \right)}}{\gamma }} \right]^4}
 + 6\alpha Sech{\left[ {\frac{{6{H^2}\left( { - 1 + q} \right)}}{\gamma }} \right]^2}\tanh {\left[ {\frac{{6{H^2}\left( { - 1 + q} \right)}}{\gamma }} \right]^2}) + j(101\beta {\gamma ^2} \\\nonumber
 &+ 48\alpha Sech{\left[ {\frac{{6{H^2}\left( { - 1 + q} \right)}}{\gamma }} \right]^4}
 - 96\alpha Sech{\left[ {\frac{{6{H^2}\left( { - 1 + q} \right)}}{\gamma }} \right]^2}\tanh {\left[ {\frac{{6{H^2}\left( { - 1 + q} \right)}}{\gamma }} \right]^2} + q(\beta {\gamma ^2} + 24\alpha\\\nonumber
 &\times Sech{\left[ {\frac{{6{H^2}\left( { - 1 + q} \right)}}{\gamma }} \right]^4}
 - 48\alpha Sech{\left[ {\frac{{6{H^2}\left( { - 1 + q} \right)}}{\gamma }} \right]^2}\tanh {\left[ {\frac{{6{H^2}\left( { - 1 + q} \right)}}{\gamma }} \right]^2}))\}]
\end{align}
For this $f(R,\Box R)$ model, we consider $0<\gamma<5,~-1<\alpha<1$
and $-1<\beta<1$ ranges of parameters and we will restrict these values by ECs. In order to obtain $\rho>0$,
one requires $\gamma<3$ along with $\alpha<0.7$ and $\beta>-0.4$. For $\rho + p >0$, the viable regions are being shown
in Fig.(\ref{f11}) for all possible value of $\gamma$ and $\alpha$ with $\beta>0$.
Furthermore, we analyze the behavior of $\rho$ and $\rho + p$ by fixing one parameter value $\gamma$, as shown in Fig.(\ref{f12}), in which we can see the positivity of both $\rho$ and $\rho + p$ with all ranges of $\gamma$ and positive values of $\beta$ in given ranges.
Similar behavior has been observed and shown in Figs.(\ref{f13}) and (\ref{f14}), in which we plot $\rho$ and $\rho + p$ with respect to $\gamma$ along with the positive as well negative values of $\alpha$. Odintsov and Oikonomou \cite{bb} presented observational consistent inflationary
constraints by providing a comparison between non-singular and singular $R^2$ cosmic model. However,
consequences of $R^2$ model consistent with latest Planck data are provided by Odintsov {et al.} \cite{ee}. Bhatti
\cite{sol1} and Yousaf \cite{sol2} provided some viable models for $\Lambda$-dominated epochs.


\subsection{Model 2}

Next, we take $f(R,\Box R,T)$ model of the form
\begin{equation}\label{model2}
f(R,\Box R) = R + \alpha \gamma \left[ {{{\left( {1 + \left( {\frac{{{R^2}}}{{{\gamma ^2}}}} \right)} \right)}^{ - \lambda }} - 1} \right] + \beta R \Box R,
\end{equation}
in which $\alpha,~\lambda$ and $\gamma$ are constants. By making use of this model and Eqs.\eqref{hq}, \eqref{jsl} and (\ref{ro}), we found
\begin{align}\nonumber
\rho  &= 18\beta {H^6}\left( { - 2 + j - q} \right)\left( {5 + j + q} \right) + \frac{1}{2}(\alpha \gamma \left( {1 - {{\left( {\frac{{{\gamma ^2} + 36{H^4}{{\left( { - 1 + q} \right)}^2}}}{{{\gamma ^2}}}} \right)}^{ - \lambda }}} \right) - 6{H^2}( - 1 + q)\\\nonumber
& + 36\beta {H^6}( - 1 + q)\left( {12 - 3j + 11q + {q^2} - s} \right)) - 18\beta {H^6}\left( {1 + 3q} \right)(6 +8q + {q^2} - s) + 6\beta {H^5} (24 - l \\\nonumber
&+ 48q + 18{q^2} + 2j(4 + q) - s) - 36\beta {H^6}\left( { - 1 + q} \right)( - 8 +
j - 11q - 3{q^2} + s) - 3{H^2}q[1\\\nonumber
 &-\left\{{{12\alpha \lambda {H^2}{{\left( {\frac{{{\gamma ^2} + 36{H^4}{{\left( { - 1 + q} \right)}^2}}}{{{\gamma ^2}}}} \right)}^{ - 1 - \lambda }}\left( { - 1 + q} \right)}}\right\}{\gamma }^{-1}+ 6\beta {H^4}(- 12 + 3j - 11q - {q^2} + s)] \\\nonumber
& + 3H[\left\{{{12\alpha \lambda {H^3}{{\left( {\frac{{{\gamma ^2} + 36{H^4}{{\left( { - 1 + q} \right)}^2}}}{{{\gamma ^2}}}} \right)}^{ - 1 - \lambda }}\left( { - 2 + j - q} \right)}}\right\}{\gamma }^{-1}\left\{864\alpha \gamma \lambda (1 + \lambda ){H^7}\right.\\\nonumber
& + \left.{\left( {\frac{{{\gamma ^2} + 36{H^4}{{\left( { - 1 + q} \right)}^2}}}{{{\gamma ^2}}}} \right)}^{ - \lambda }{{\left( { - 1 + q} \right)}^2}\left( {2 - j + q} \right)\right\}....
\left\{{{{\left( {{\gamma ^2} + 36H{{\left[ t \right]}^4}{{\left( { - 1 + q\left[ t \right]} \right)}^2}} \right)}^2}}\right\}^{-1}\\\nonumber
& - 6\beta {H^5}\left( { - 48 + l + j\left( { - 5 + q} \right) - 81q - 24{q^2} + 4s} \right)],
\end{align}
while Eqs.(\ref{ro}) and (\ref{ropp}) yield
\begin{align}
\rho+ p  &= 18\beta {H^6}\left( { - 2 + j - q} \right)\left( {5 + j + q} \right) + 6\beta {H^6}\left( { - 2 + j - q} \right)(29 + 4j
 + 10q) -\left\{{{\left( {{\gamma ^2} + 36{H^4}{{\left( { - 1 + q} \right)}^2}} \right)}^3}\right\}^{-1}\\\nonumber
& \times12\alpha \gamma \lambda {H^4}{\left( {\frac{{{\gamma ^2} + 36{H^4}{{\left( { - 1 + q} \right)}^2}}}{{{\gamma ^2}}}} \right)^{ - \lambda }}(144(1
 + \lambda ){H^4}\left( {{\gamma ^2} + 36{H^4}{{\left( { - 1 + q} \right)}^2}} \right)\left( { - 1 + q} \right){\left( {2 - j + q} \right)^2}\\\nonumber
 & - 72\left( { - 1 - \lambda } \right){H^4}\left( { - 1 + q} \right)
[({\gamma ^2} + 36{H^4}( - 1{\left( { - 1 + q} \right)^2}){\left( {2 - j + q} \right)^2} - 72\left( {2 + \lambda } \right){H^4}{\left( { - 1 + q} \right)^2}(2 + q\\\nonumber
& - j )^2- \left( {{\gamma ^2} + 36{H^4}{{\left( { - 1 + q} \right)}^2}} \right)\left( { - 1 + q} \right)(6 + 8q + {q^2} - s)) + {\left( {{\gamma ^2} + 36{H^4}{{\left( { - 1 + q} \right)}^2}} \right)^2}
(6 + 8q + {q^2}\\\nonumber
&- s)]- 18\beta {H^6}( 1 + 3q)\left( {6 + 8q + {q^2} - s} \right) - 6\beta {H^6}(13
 + 5q)\left( {6 + 8q + {q^2} - s} \right) + 6\beta {H^5}( 24 - l + 48q \\\nonumber
 & + 18{q^2}+ 2j(4 + q) - s )
 + 30\beta {H^6}\left( {24 - l + 48q + 18{q^2} + 2j\left( {4 + q} \right) - s} \right)
 - 48\beta {H^6}\left( { - 1 + q} \right) ( - 8 + j\\\nonumber
& - 11q - 3{q^2} + s) + 6\beta {H^6}(240 - {j^2} + 7l + m + 564q
 + 315{q^2} + 24{q^3} + 8j\left( {12 + 7q} \right) + 57s + 25qs) \\\nonumber
 &- 6\beta {H^6}(120 + 2{j^2} + 10l + m + 312q
 + 6\beta {H^4}\left( { - 12 + 3j - 11q - {q^2} + s} \right)) - 3{H^2}q(1- \gamma^{-1}\\\nonumber
&  \times\left\{{12\alpha \lambda {H^2}{{\left( {\frac{{{\gamma ^2} + 36{H^4}{{\left( { - 1 + q} \right)}^2}}}
  {{{\gamma ^2}}}} \right)}^{ - 1 - \lambda }}\left( { - 1 + q} \right)}\right\} + 6\beta {H^4}( - 12 + 3j - 11q
 - {q^2} + s)) + H\\\nonumber
 &\times(\left\{{12\alpha \lambda {H^3}{{\left( {\frac{{{\gamma ^2} + 36{H^4}{{\left( { - 1 + q} \right)}^2}}}{{{\gamma ^2}}}} \right)}^{ - 1 - \lambda }}\left( { - 2 + j - q} \right)}\right\}{\gamma }^{-1}
 + 864\alpha \gamma \lambda (1 + \lambda ){H^7} \left( {\gamma ^2} + 36{H^4}\right.\\\nonumber
&\times\left.{{\left( { - 1 + q} \right)}^2}{\gamma ^2}^{-1} \right)^{ - \lambda }{{\left( { - 1 + q} \right)}^2}
\left( {2 - j + q} \right)\left\{{{{\left( {{\gamma ^2} + 36H{{\left[ t \right]}^4}{{\left( { - 1 + q\left[ t \right]} \right)}^2}} \right)}^2}}\right\}^{-1}
 - 6\beta {H^5}( - 48 + l + j\\\nonumber
 & \times \left( { - 5 + q} \right) - 81q - 24{q^2} + 4s ))
\end{align}

For our second $f(R,\Box R,T)$ model, we consider the values of $\gamma,~\alpha,~\lambda$ and $\beta$ from the closed intervals
$[0,~5],~[-1,~1],~[-1,~1],~$ and $[-1,~1]$, respectively. We shall put constraints on these parameters via ECs.
\begin{figure} \centering
\epsfig{file=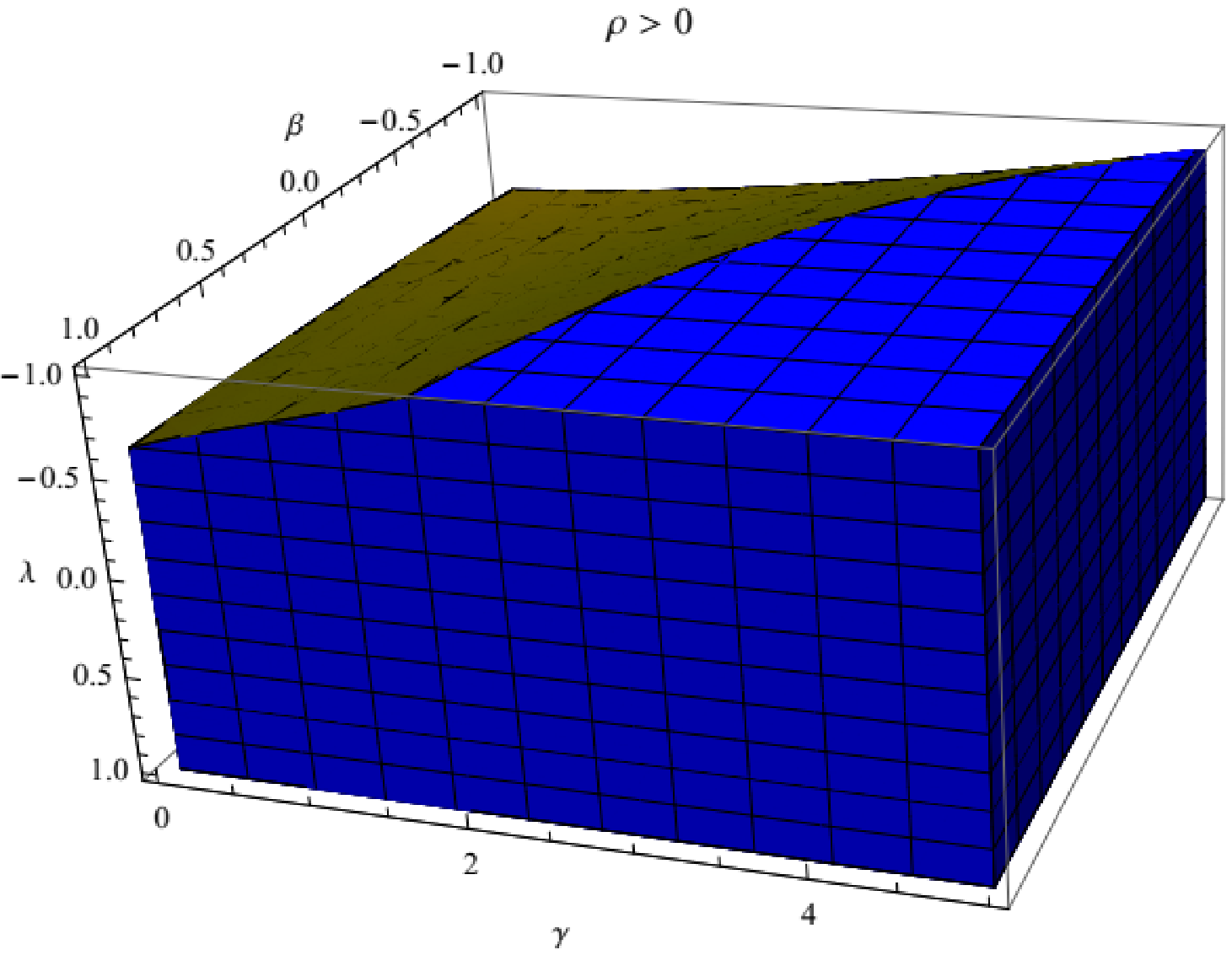,width=.48\linewidth}
\epsfig{file=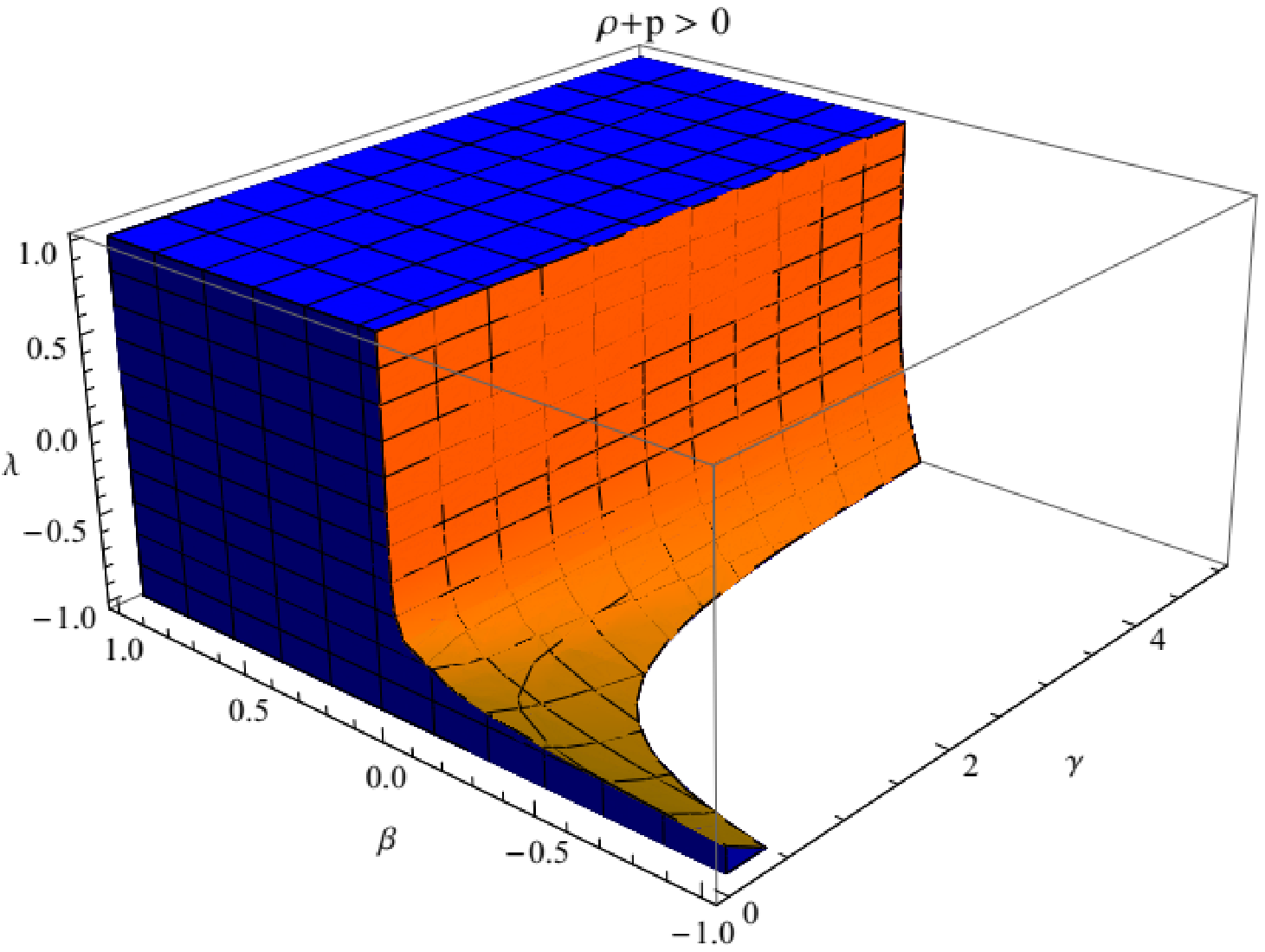,width=.48\linewidth}
\caption{Plot of WEC for model 2 as in Eq.(\ref{model2}), left plot show $\rho$ while the right plot show the $\rho+ p$ with respect to $\lambda$, $\gamma$ and $\beta$}\label{f21}
\end{figure}
We noticed that $\rho>0$ requires $\lambda>-0.6$, while $\rho + p>0$ requires positive values of $\beta$. There are also small regions for which $\rho + p >0$. The details can be observed from Fig.(\ref{f21}).
\begin{figure} \centering
\epsfig{file=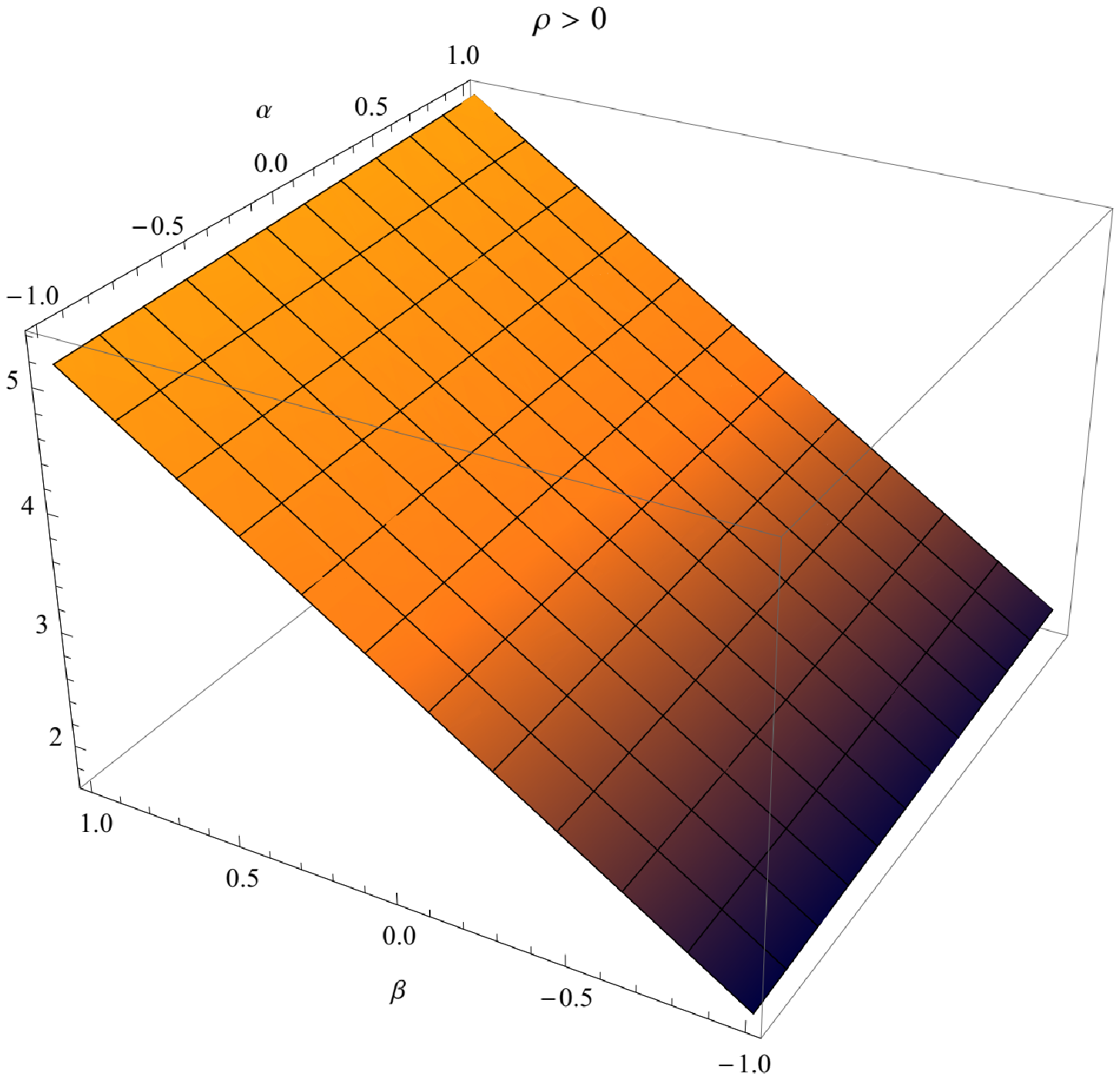,width=.48\linewidth}
\epsfig{file=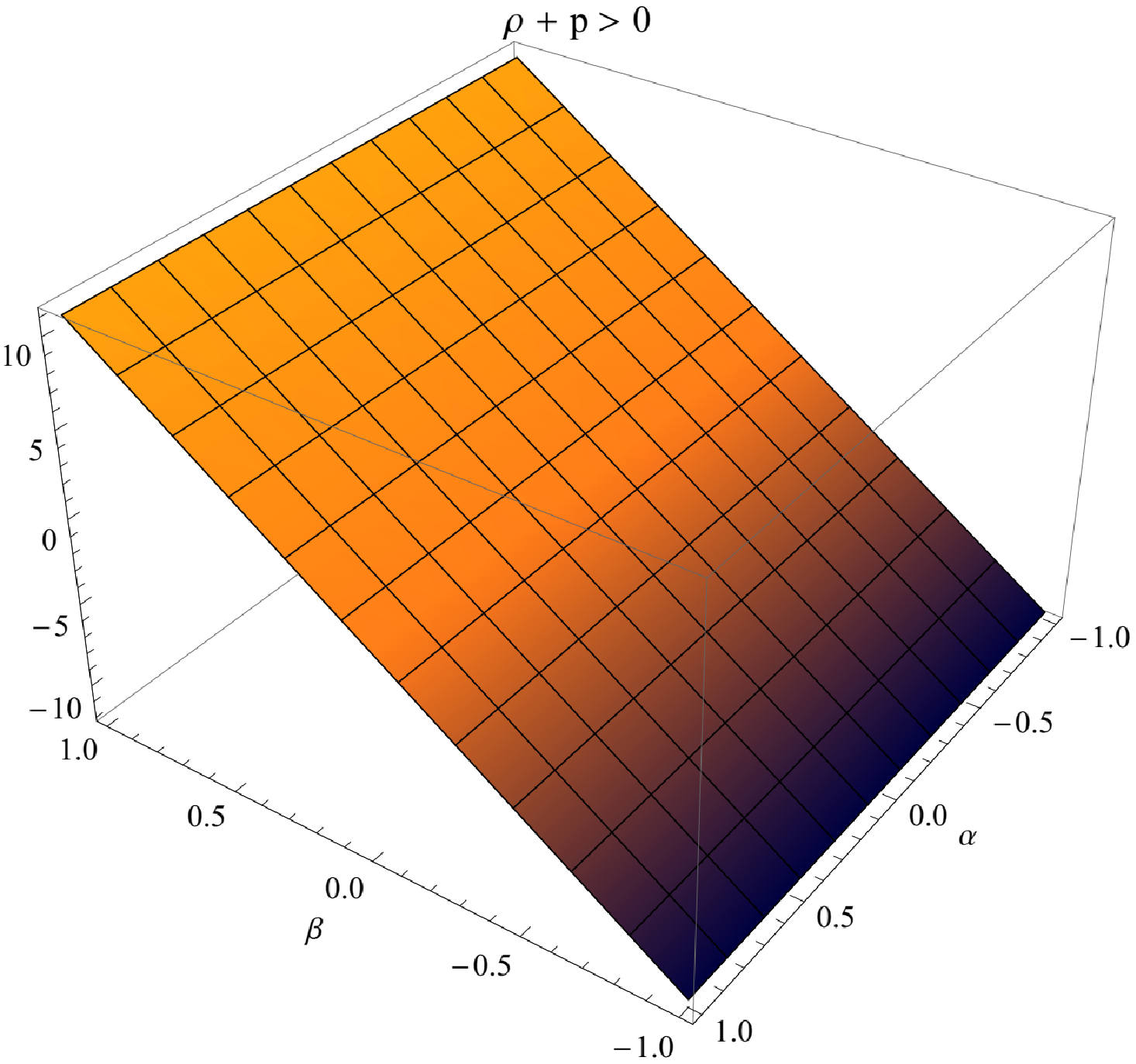,width=.48\linewidth}
\caption{Plot of WEC for model 2 as in Eq.(\ref{model2}), left plot show $\rho$ while the right plot show the $\rho+ p$ with respect to $\alpha$ and $\beta$}\label{f22}
\end{figure}
By fixing one both $\gamma$ and $\lambda$ values, we plotted the viable regions of ECs as Fig.(\ref{f22}). One can observed the positive behavior of $\rho$ for all given ranges of $\alpha$ and $\beta$, while $\rho + p>0$ requires $\beta$ to be greater than zero.
\begin{figure} \centering
\epsfig{file=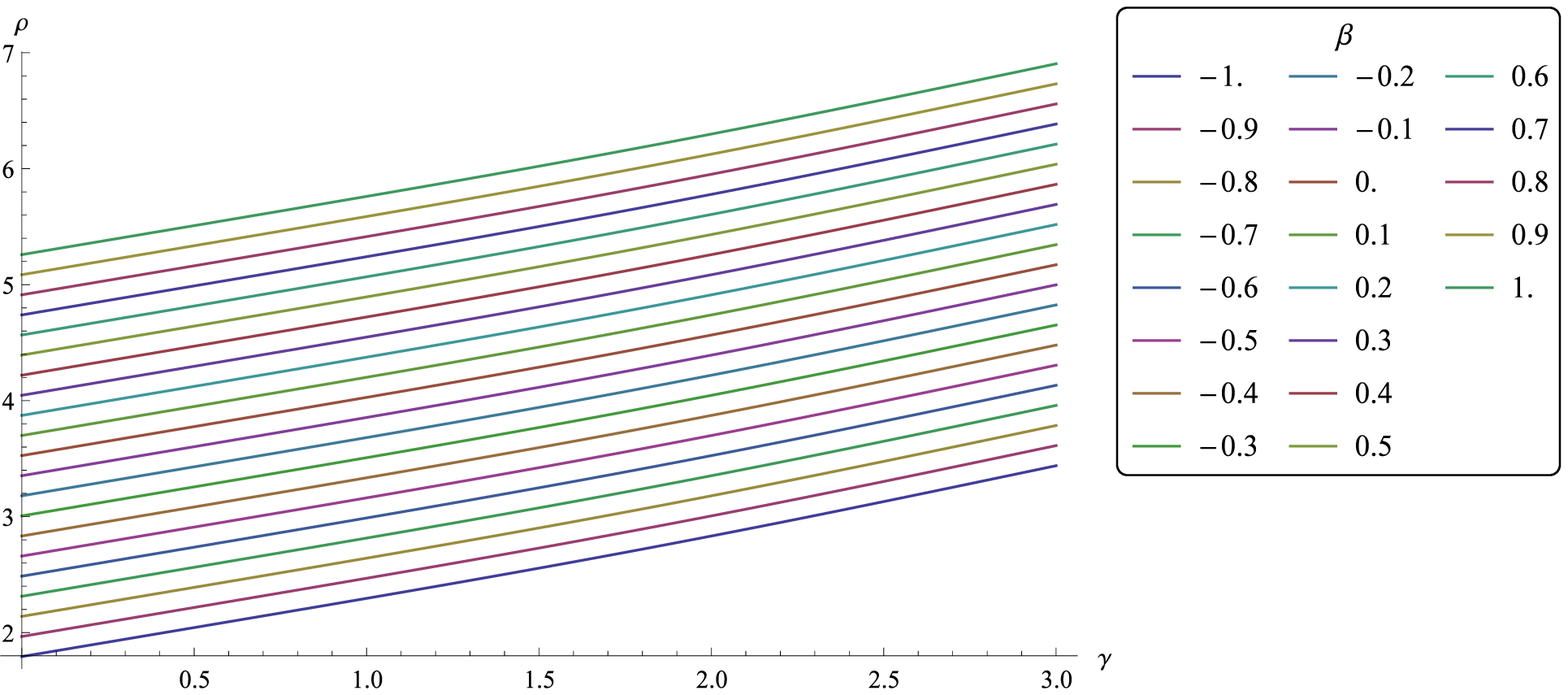,width=.48\linewidth}
\epsfig{file=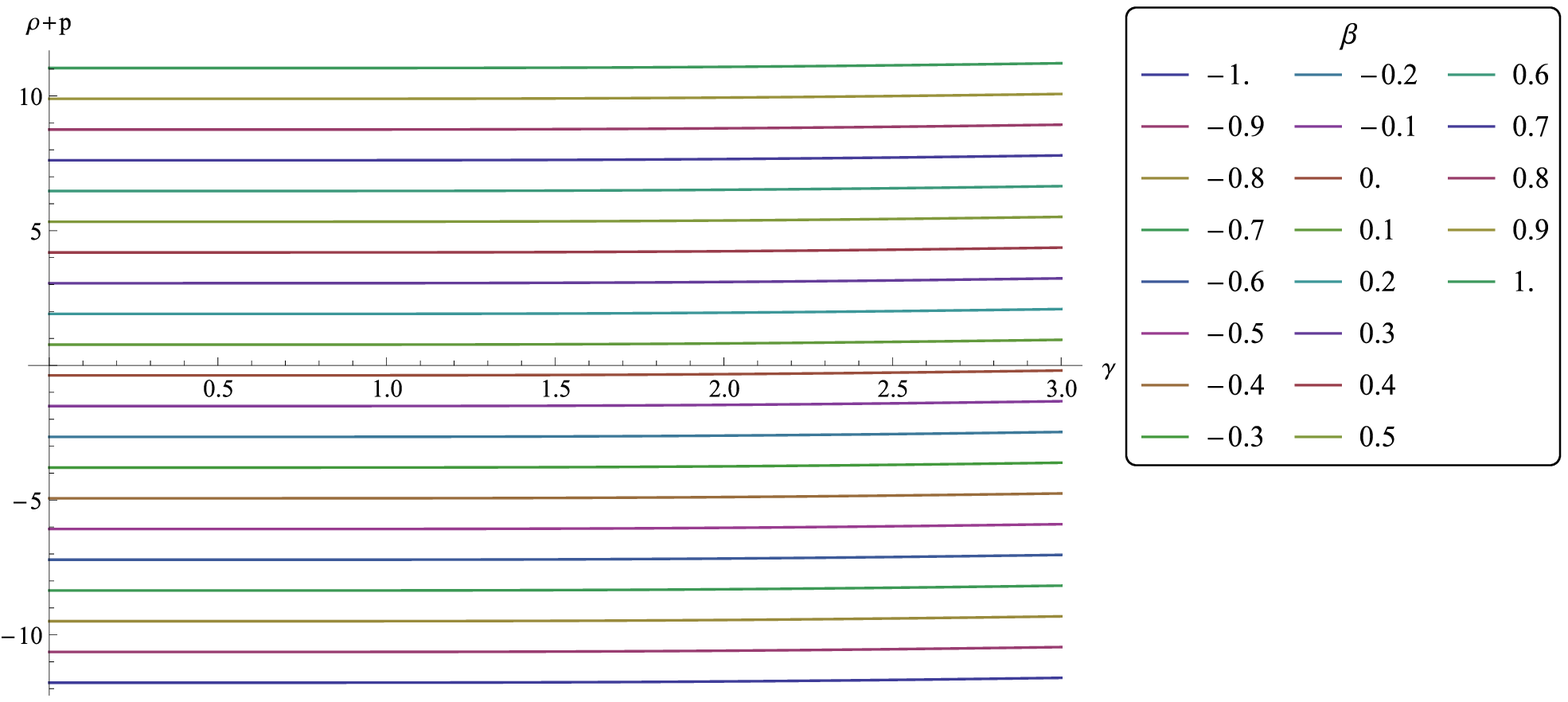,width=.48\linewidth}
\caption{Plot of WEC for model 2 as in Eq.(\ref{model2}), left plot show $\rho$ while the right plot show the $\rho+ p$ with respect to $\gamma$ with $\alpha>0$.}\label{f23}
\end{figure}
\begin{figure} \centering
\epsfig{file=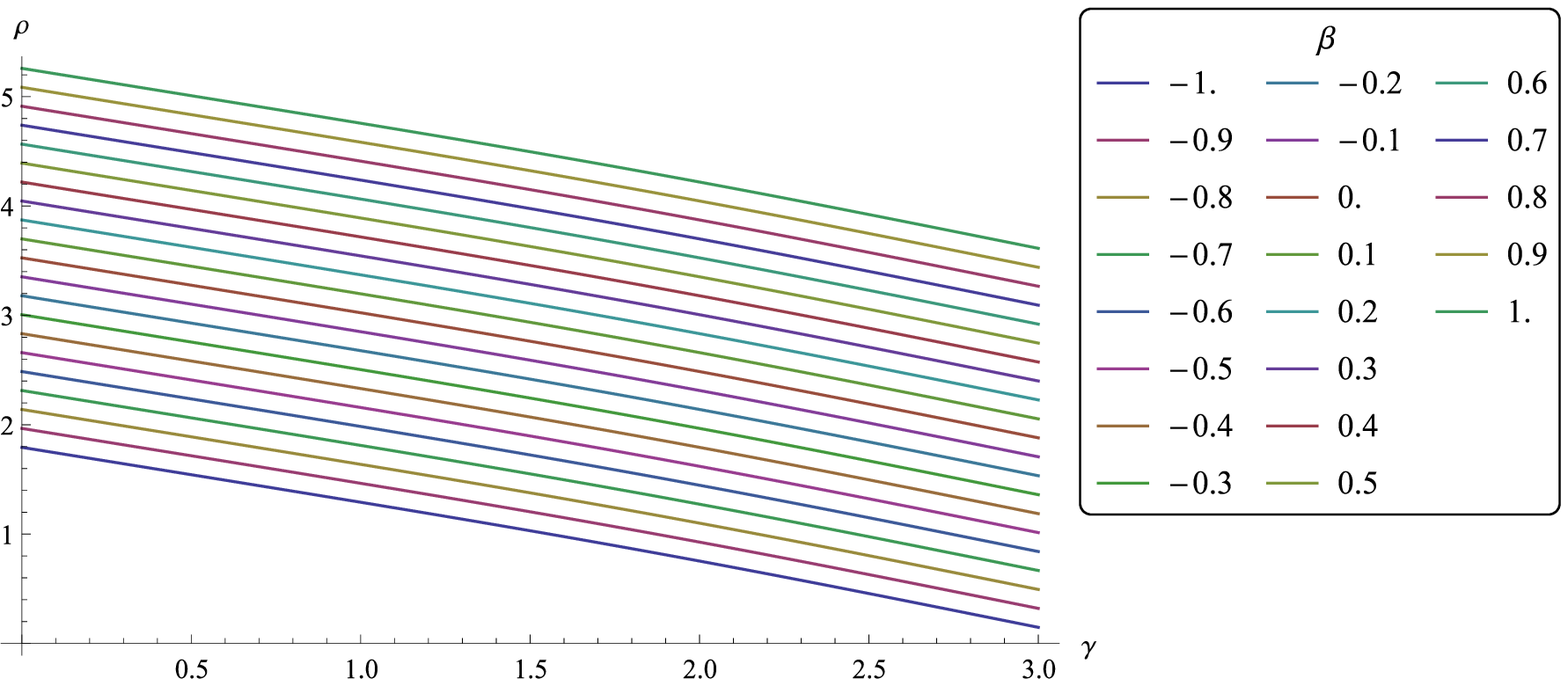,width=.48\linewidth}
\epsfig{file=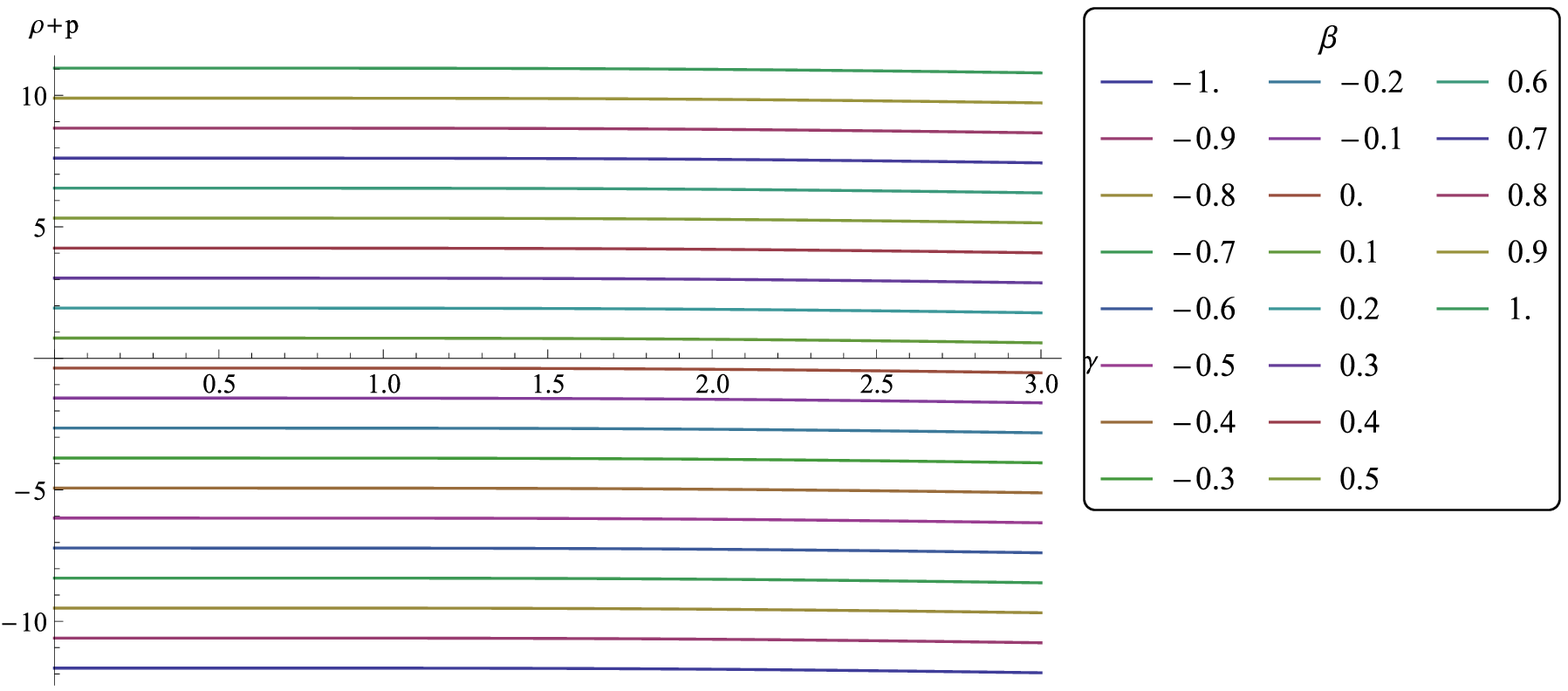,width=.48\linewidth}
\caption{Plot of WEC for model 2 as in Eq.(\ref{model2}), left plot show $\rho$ while the right plot show the $\rho+ p$ with respect to $\gamma$ with $\alpha<0$.}\label{f24}
\end{figure}
Now, by taking different values of $\beta$ with $\alpha>0$, we have plotted the diagrams for $\rho$ and $\rho+ p$ as shown in Fig.(\ref{f23}). We conclude that for the validity of ECs, $\beta$ should be greater than zero. We check the similar behavior for $\alpha<0$, which gives the same conclusion e.g. $\beta$ should be non-negative real number. The details of these can be seen from Fig.(\ref{f24}).

\section{Summary}

The modified gravity theories have been emerged as among good candidates to study the cosmic acceleration
of the expanding universe. The $f(R,\Box R,T)$ gravity theory have gained significance on the basis of curvature matter coupling
and can be considered as a generalization of $f(R,T)$ gravity theory. However, there is a crucial difference
between both modified theories due to the higher derivative terms of the Ricci scalar in the gravitational Lagrangian and consequently leads to a significant deviation from the geodesic paths.
In this paper, we have discussed the ECs in the context of $f(R,\Box R,T)$ gravity theory with two different
models which is the best viable method to test the validity of these theories.
The WEC has been evaluated using the Raychaudhuri
equation which are more general as compared to those obtained in $f(R)$ and $f(R,T)$ gravity theories. We showed that these energy conditions can be satisfied with the modified gravity models. We have found that the obtained inequalities have equivalence
with those obtained via $p+\rho\geq0$ and $\rho\geq0$ with the limit $p\rightarrow p^{eff}$ and $\rho\rightarrow\rho^{eff}$.
We have considered two functional forms of $f$ namely $R-\alpha\gamma\tanh\left({\frac{R}{\gamma}}\right)+\beta R\Box R$ and
$R+\alpha\gamma\left[{{{\left({1+\left({\frac{{{R^2}}}{{{\gamma^2}}}}\right)}\right)}^{-\lambda}}-1}\right]+\beta R\Box R$.
We have used the recent estimated values of the snap, deceleration, jerk and Hubble parameters to show the validity of the different functional forms of $f(R,\Box R,T)$ imposed by the WEC.

For the first functional value of the $f(R,\Box R,T)$ model, we found particular constraints on the parameters $\alpha,~\beta$ and $\gamma$ to satisfy the energy conditions. We observed the positivity of energy density for the first model when $\alpha<0.7,~\beta>-0.4$ and $\gamma<3$, however, the positivity of $\rho+p>0$ is observed when $\beta>0$ with all possible values of $\alpha$ and $\gamma$. These results are indicated in Fig. (\ref{f11}). Similar results have been obtained by fixing the parametric value of $\gamma$ and with different values of $\alpha$ and $\beta$ as shown in Fig. (\ref{f12}). We have also discussed the positivity of energy density and $\rho+p$ for different values of $\beta$ with positive and negative $\alpha$ and presented via plots in Figs.(\ref{f13}) and (\ref{f14}). Similarly, for the second functional value of the $f(R,\Box R,T)$ model, we found particular constraints on the parameters $\alpha,~\beta,~\gamma$ and $\lambda$ to satisfy the energy conditions. We have observed that $\rho>0$ for the second model when $\lambda>-0.6$, however, the positivity of $\rho+p>0$ is observed when $\beta>0$ with all possible values of $\alpha$ and $\gamma$. These results are indicated in Fig. (\ref{f21}). Similar results have been obtained by fixing the parametric value of $\lambda$ and $\gamma$ and plotted with respect to $\alpha$ and $\beta$ as shown in Fig. (\ref{f22}). We have also discussed the positivity of energy density and $\rho+p$ for different values of $\beta$ with positive and negative $\alpha$ and presented via plots in Figs.(\ref{f23})-(\ref{f24}). One can conclude that the viability of ECs depend on particular values of the model parameters.

It is significant to mention here that regardless of well-motivated physical interpretation of ECs in modified gravity theories, it is still under discussion due to the confrontation between the observations and the theory. We stress that our method and interpretation is rather general
from the context of higher derivative theory which can be reduced to other modified gravity results under the usual limits.

\end{document}